# Educational approach to general relativity through its approximate vectorial form

Kjell Prytz

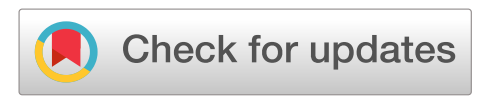

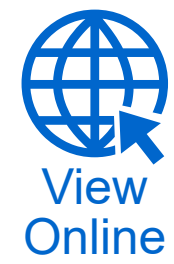

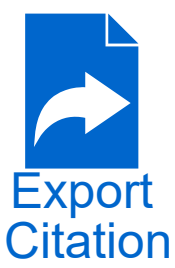

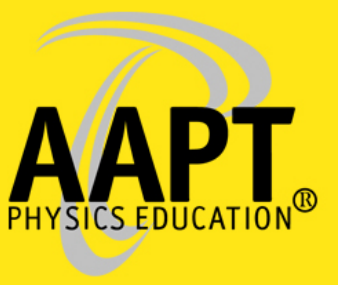

# Educational approach to general relativity through its approximate vectorial form

Kjell Prytz[a)]
*Malardalen University, Vasteras, Sweden*

General Relativity (GR) can be effectively examined in the low-velocity, weak-field limit, where its inherently tensorial structure simplifies to a vector-based formulation closely analogous to classical electrodynamics. This correspondence becomes particularly clear when both theories are expressed within a field-free, interaction-based framework, in which the leading relativistic corrections emerge naturally as velocity- and acceleration-dependent modifications to the static interaction. As a result, several predictions of GR can be derived using familiar techniques from electrodynamics, provided the relevant phenomena remain within the regime of validity of the approximation. This structural analogy facilitates a conceptual understanding of GR enabling the introduction of its core ideas even in an introductory level course, the single exception being the treatment of radiation, which requires tools at more advanced level. Building on this pedagogical framework, the incorporation of special relativity (SR) and the equivalence principle enables concise and transparent derivations of gravitational time dilation, as well as its immediate consequences, including light deflection and gravitational frequency shifts. A detailed exposition of this material is provided in the supplementary material. 

## I. INTRODUCTION

The objective of this paper is to present an introductory approach to General Relativity (GR) that circumvents the technical complexity of tensor calculus and differential geometry. Instead, the focus is placed on the weak-field, low-velocity limit of the theory, where a compelling formal analogy with classical electromagnetism emerges.

Two levels of approximation are commonly employed in general relativity. The first involves the weak-field limit, wherein gravitational interactions are sufficiently weak to justify neglecting the nonlinearities arising from the self-interacting nature of the gravitational field. This permits the preservation of the full tensorial framework of general relativity, though in a linearized form, thereby simplifying the search for analytical solutions. The second level of approximation proceeds by assuming that the velocities of the interacting bodies are much smaller than the speed of light. Under this condition, the tensorial structure of the theory effectively reduces to a vectorial form.

The formulation adopted in this work is based on Einstein's original weak-field, low-velocity approximation,[1] which was initially cast in terms of gravitational potentials but can be equivalently expressed using field quantities analogous to electric and magnetic fields, a framework commonly referred to as Gravito-Electromagnetism.[2] In this paper, we present an interaction-based approach that provides an alternative pedagogical perspective, building upon and extending recent developments.[3] This approach is field-free and explicitly models the complete interaction between physical objects, rather than isolating contributions through intermediary fields.

By embracing an interaction-based approach, it becomes straightforward to identify corrections to Newton's law of gravitation arising from uniform and accelerated motion—mirroring the modifications to Coulomb's law in electrodynamics. By emphasizing the structural parallels between GR and electrodynamics, this method opens the door to teaching GR conceptually, grounded in observations rather than abstract mathematics. Because magnetic and inductive effects, familiar from high school-level physics, have analogues in GR, learners can draw on well-known physical intuition.

In this paper, we begin by formulating classical electrodynamics within the interaction picture. We then show that a parallel formalism naturally arises from the linearized Einstein field equations in the low-velocity regime. This framework is subsequently applied to derive several foundational predictions of General Relativity (GR) and to illuminate the physical mechanisms underlying these phenomena. As a pedagogical extension, we demonstrate how Special Relativity (SR), when combined with the equivalence principle, provides an intuitive pathway for deriving gravitational time dilation and related effects. This part is provided as a supplementary material. Finally, within the same approximation scheme, we outline the derivation of gravitational radiation power from a field-theoretic perspective, highlighting its close analogy to electromagnetic radiation. Notably, all these results are obtained within the context of flat spacetime, without invoking the conventional geometrical approach to GR through spacetime curvature. Apart from the treatment of radiation, all results are field-free and based on forces in flat spacetime.

## II. KINEMATICAL EFFECTS IN ELECTROMAGNETISM

The kinematical effects in classical electrodynamics are primarily derived based on current-carrying conductors, where the velocities of the charges are typically low. The foundational work for these phenomena was laid by

  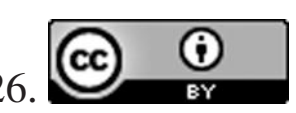

Ampère[4] and Faraday,[5] who investigated the magnetic and inductive phenomena, respectively. These effects correspond to the velocity- and acceleration-dependent dynamics in electrodynamics. Typically, textbooks present these results through Maxwell's equations, which provide a comprehensive field-theoretical description of electromagnetism.

Weber was the first to fully formulate classical electrodynamics in terms of interactions between elementary charged constituents.[6] In modern terms, this force-based approach was re-expressed by Moon and Spencer in the 1950s,[7] and may be written as

$$\bar{f}_{1\to 2} = \bar{f}_{\text{static}} + \bar{f}_{\text{magn}} + \bar{f}_{\text{ind}} = \frac{q_1 q_2}{4\pi\varepsilon_0 R^2}\hat{R} + \frac{q_1 q_2}{4\pi\varepsilon_0 c^2 R^2}\Big((-\bar{v}_1\cdot\bar{v}_2)\hat{R} + (\bar{v}_2\cdot\hat{R})\bar{v}_1\Big) - \frac{q_2 q_1}{4\pi\varepsilon_0 c^2 R}\frac{d\bar{v}_1}{dt}. \tag{1}$$

This equation represents the force on object 2 in an interaction with object 1. The vector $\bar{R} = \bar{r}_2 - \bar{r}_1$ points from object 1 to object 2 and $\hat{R}$ is its unit vector (see Fig. 1). The first term is the electrostatic force, the second and third terms account for the magnetic force, and the fourth term represents the inductive force.

In electrodynamics, the magnetic and inductive forces between free charges are generally too small to be directly measurable. Consequently, empirical data for these forces are obtained primarily from interactions involving current-carrying closed conductors. Additional evidence is available from experiments involving interactions between a free charged particle beam and a closed conductor. Formula (1) represents a two-particle interaction inferred from such experimental configurations. It follows that the magnetic term is physically meaningful only when object 1 forms part of a closed conductor. Furthermore, if object 2 is also part of a closed conductor, the second magnetic term vanishes. This result arises from symmetry considerations associated with the closed current path. Detailed derivations of this property are provided in Ref. 11, Chap. 2.

The inductive force is deduced exclusively from experiments involving closed conductors, which consequently define the domain of its applicability. It manifests as a reactive force opposing changes in electric current, thereby giving rise to the phenomena of inductance and magnetic energy storage.

The magnetic and inductive terms in electrodynamics are fundamentally kinematical effects, derivable from Lorentz transformations.[8] As such, special relativity provides a unified *kinematical* account of magnetism and electromagnetic induction, showing them to be frame-dependent aspects of a single electromagnetic interaction—an achievement which, in my view, is arguably its most profound.

## III. GENERAL RELATIVITY IN VECTORIAL REPRESENTATION

The kinematical effects in electrodynamics are a consequence of the fact that interactions propagate at a finite speed. By analogy, if gravitational interactions also propagate at a finite speed, as is the case in general relativity, similar velocity- and acceleration-dependent effects are anticipated. Therefore, analogous to Eq. (1), the following interaction-based equation can be hypothesized for the gravitational force acting on object 2, due to object 1:

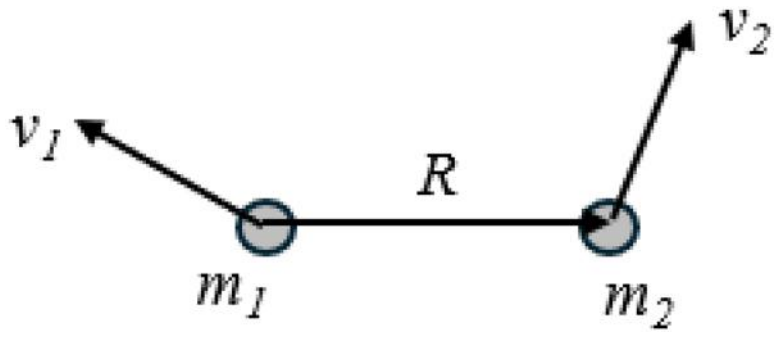

Fig. 1. Kinematics for calculating the force on object 2 due to object 1.

$$\bar{f}^{g}_{1\to 2} = \bar{f}_{g-\text{static}} + \bar{f}_{g-\text{magn}} + \bar{f}_{g-\text{ind}} = -\frac{Gm_1 m_2}{R^2}\hat{R} - k_m\frac{Gm_1 m_2}{c^2R^2} \times \Big((-\bar{v}_1\cdot\bar{v}_2)\hat{R} + (\bar{v}_2\cdot\hat{R})\bar{v}_1\Big) + k_i\frac{Gm_1 m_2}{c^2 R}\frac{d\bar{v}_1}{dt}. \tag{2}$$

The coefficients $k_m$ and $k_i$ are constants that will be determined from GR. The first term represents the static Newtonian gravitational force. The second and third terms depend on velocity and are called gravitomagnetic terms. The fourth term accounts for the gravitational acceleration-dependent interaction, so-called gravito-induction. The negative sign in the static term reflects the inherently attractive nature of gravity, in contrast to the repulsive interaction between like charges in electromagnetism. This sign reversal propagates to the other terms.

Formula (2) is now compared to the expression given by Einstein in the same approximation scheme which is up to order $v^2/c^2$. Apart from a factor on the left-hand side, considered later, Eq. (118) in Ref. 1 reads

$$\frac{d}{dt}\frac{d\bar{r}_2}{dt} = -\frac{Gm_1}{R^2}\hat{R} + c\big(\nabla_2\times\bar{A}_1\big)\times\frac{d\bar{r}_2}{dt} + c\frac{d\bar{A}_1}{dt}, \tag{3}$$

which is the acceleration of object 2 caused by object 1 at a distance $R$.

To obtain Eq. (3) from Eq. (188) in Ref. 1, we have proceeded as follows. Einstein′s notations are as follows: $dl = cdt$, $\bar{v} = d\bar{r}_2/dl$, $\sigma$ is mass density, $\kappa = 8\pi G/c^2$, and $rot\,\bar{A} = \nabla\times\bar{A}$.

We may then write

$$\bar{A}_1 = \frac{4Gm_1}{c^2R}\frac{1}{c}\frac{d\bar{r}_1}{dt}, \tag{4}$$

which is the gravitational vector potential due to object 1 [Ref. 1, Eq. (118)], equivalent to the electrodynamical vector potential. Thus, $c$ times the curl of Eq. (4) is the gravitomagnetic field. It should be evaluated at the coordinates of object 2 and acts therefore only on position vector $\bar{r}_2$, where $\bar{R} = \bar{r}_2 - \bar{r}_1$, so

$$\nabla_2\times\frac{d\bar{r}_1}{R} = \frac{d\bar{r}_1\times\hat{R}}{R^2}. \tag{5}$$

The second term in Eq. (3) can now be written as follows:

$$c\,\frac{4Gm_1}{c^3}\frac{\frac{d\bar{r}_1}{dt}\times\hat{R}}{R^2}\times\frac{d\bar{r}_2}{dt}$$
$$=\frac{4Gm_1}{c^2}\frac{\bar{v}_1\times\hat{R}}{R^2}\times\bar{v}_2$$
$$=-\frac{4Gm_1}{c^2R^2}\left(\bar{v}_1\left(\bar{v}_2\cdot\hat{R}\right)-\hat{R}\left(\bar{v}_1\cdot\bar{v}_2\right)\right). \tag{6}$$

The force on object 2 then becomes

$$\bar{f}_{1\to 2}=m_2\frac{d\bar{v}_2}{dt}$$
$$=-\frac{Gm_2m_1}{R^2}\hat{R}-\frac{4Gm_2m_1}{c^2R^2}$$
$$\times\left(-\hat{R}\left(\bar{v}_1\cdot\bar{v}_2\right)+\left(\bar{v}_2\cdot\hat{R}\right)\bar{v}_1\right)$$
$$+\frac{4Gm_2m_1}{c^2R}\frac{d\bar{v}_1}{dt}, \tag{7}$$

which equals formula (2) if $k_m = k_i = 4$. Thus, as expected, the kinematical effects appear equivalently in GR and in electrodynamics, apart from the factor of 4 which, as we will see, is crucial in experimental tests.

However, this almost direct transcription from electrodynamics to gravitation omits an important aspect: the relativistic modification of inertia. In general relativity, Einstein derived such an effect in the weak-field limit from the full nonlinear theory. He argued that, according to GR, the inertial mass of a body increases in the presence of a gravitational field of another mass. When this additional effect is included, the full expression given by Einstein [Ref. 1, Eq. (118)] reads

$$f^c_{1\to 2}=m_2\left(1+\frac{Gm_1}{c^2R}\right)\frac{d\bar{v}_2}{dt}$$
$$=-\frac{Gm_2m_1}{R^2}\hat{R}-\frac{4Gm_2m_1}{c^2R^2}$$
$$\times\left(-\hat{R}\left(\bar{v}_1\cdot\bar{v}_2\right)+\left(\bar{v}_2\cdot\hat{R}\right)\bar{v}_1\right)$$
$$+\frac{4Gm_2m_1}{c^2R}\frac{d\bar{v}_1}{dt}. \tag{8}$$

The correction on the left-hand side, denoted by upper index $c$, implies that the inertial mass is effectively increased by the gravitational interaction energy divided by $c^2$, consistent with the mass–energy equivalence relation $E = mc^2$ from special relativity. In the present context, we are concerned with the weak-field and low-speed limit, retaining terms up to order $1/c^2$ in the force formula. To make this explicit, we divide both sides of Eq. (8) by the multiplicative correction factor $1 + (Gm_1/c^2R)$ and then apply a Taylor expansion. Keeping terms only to first order in $1/c^2$ yields

$$\bar{f}_{1\to 2}=m_2\frac{d\bar{v}_2}{dt}$$
$$=-\frac{Gm_2m_1}{R^2}\left(1-\frac{Gm_1}{c^2R}\right)\hat{R}-\frac{4Gm_2m_1}{c^2R^2}$$
$$\times\left(-\hat{R}\left(\bar{v}_1\cdot\bar{v}_2\right)+\left(\bar{v}_2\cdot\hat{R}\right)\bar{v}_1\right)$$
$$+\frac{4Gm_2m_1}{c^2R}\frac{d\bar{v}_1}{dt}. \tag{9}$$

Thus, even in a purely static scenario, a GR correction to Newton's law of gravitation arises due to the gravitational potential energy's contribution to inertia. This correction is second order in $G$. But if $m_1$ is large, the correction might not be negligible, as will be illustrated when analyzing perihelion shift in Sec. IV.

Note that Eq. (9) is obtained by taking Einstein's formula (118) as the starting point. This formula is derived to first order in $v^2/c^2$, corresponding to the first post-Newtonian approximation of general relativity. To obtain the next-order corrections, one must return to Einstein's original procedure in Ref. 1 and retain terms of order $v^4/c^4$.

## IV. APPLICATIONS OF GR, EQUATION (9)

### A. Parallel motion

As a first application of GR in the adopted approximation, consider two objects moving in parallel to each other at constant velocity $v$ (Fig. 2). Their velocities are equal and perpendicular to the separation vector $R$. The second magnetic term therefore vanishes. The acceleration is zero, so the last term vanishes too. The gravitational field is weak, so the term containing $G^2$ can be neglected. Equation (9) reduces to

$$\bar{f}_{1\to 2}=-\frac{Gm_2m_1}{R^2}\left(1-4\frac{v^2}{c^2}\right)\hat{R}, \tag{10}$$

where the second term represents the gravitomagnetic correction, oppositely directed to the Newtonian force. This behavior is analogous to that observed in classical electrodynamics, where two electric charges moving in parallel experience a reduced mutual force due to the magnetic force. If Eq. (10) is applied blindly, the force would reverse sign when speed $v > (\frac{1}{2})c$. The low-speed approximation disallows this conclusion. It is at present not known how this interaction appears in full GR, but is quite feasible to explore in higher-order post-Newtonian treatments or solving the full nonlinear GR equations numerically.

Although this GR effect has not been observed, it serves as a valuable illustrative example. It introduces the concept of gravitomagnetic interaction and underscores the deep conceptual parallels between general relativity and electromagnetism, while also drawing attention to a key distinction—namely, the appearance of that factor of 4 in Eq. (10).

### B. Lense–Thirring frame dragging effect

The Lense–Thirring effect deals with the rotation of a massive object and is usually expressed as a distortion of the surrounding spacetime, effectively "dragging" it along with the spinning mass.[9] In the interaction picture, this implies that a moving test particle in the vicinity of a rotating mass will experience a deflection in its trajectory within the plane of rotation. The direction of this deflection depends on the relative direction of motion between the particle and the spinning source.

To build intuition, the rotating mass can be modeled as a closed loop of circulating mass, a mass current (Fig. 3). Consider a test particle approaching this rotating system from the outside. Focus on the nearest mass element of the loop. Since the relative velocities between the test particle $m_2$ and this mass element $m_1$ of the loop are perpendicular, Eq. (9) reduces to

$$\bar{f}_{1\to 2} = -\frac{Gm_2m_1}{R^2}\left(1-\frac{Gm_1}{c^2R}\right)\hat{R} - \frac{4Gm_2m_1}{c^2R^2}\left(\bar{v}_2\cdot\hat{R}\right)\bar{v}_1 + \frac{4Gm_2m_1}{c^2R}\frac{d\bar{v}_1}{dt}. \quad (11)$$

The last term in Eq. (11) originates from the centripetal acceleration of the mass current and, accordingly, is directed radially. The middle term, in the direction of $\bar{v}_1$, represents the Lense–Thirring effect, a manifestation of gravitomagnetism. Since the scalar product in this term is negative, the resulting gravitomagnetic force is directed parallel to the mass current. This implies that an incoming particle will acquire angular momentum from the rotating mass, effectively beginning to co-rotate with it. Analogous to classical electromagnetism, this process may be interpreted as the result of a gravitomagnetic field generated by the circulating mass current. The force experienced by the approaching object can be seen as a gravitational counterpart to the Lorentz force, acting perpendicular to both the gravitomagnetic field, being directed into the plane in Fig. 3, and the velocity of the moving particle. If the particle moves outward, the direction of the bending force reverses.

A similar analysis applies to a particle situated *inside* of the circulating mass current. If the particle is not precisely at the center, it experiences an attractive force toward the nearest mass elements due to the static (Newtonian) gravitational field. Additionally, the particle undergoes a deflection in the direction of rotation because of the gravitomagnetic interaction. Einstein described this phenomenon as analogous to a particle moving within a "Coriolis field."

These fundamental manifestations of the Lense–Thirring effect, i.e., the direct deflection of particle trajectories by local gravitomagnetic fields, have not been observed. To obtain an observable effect, consider the spin–spin interaction between rotating bodies. In electrodynamics, a spinning charged object, such as an electron or atomic nucleus, undergoes Larmor precession of its angular momentum vector when placed in an external magnetic field. An analogous phenomenon is predicted in gravitation: a spinning mass in a gravitomagnetic field experiences precession of its spin axis. The direction and rate of this precession has been derived within the framework of Gravitoelectromagnetism.[10]

To determine the gravitational precession rate in our adopted interaction picture, we start from the well-known expression for the electromagnetic dipole–dipole interaction.[11] Substituting $\mu_0/4\pi \to G/c^2$, converting from gravitomagnetic dipole moment $\bar{\mu}_g$ to spin angular momentum $\bar{S}$ through the relation $\bar{\mu}_g = \bar{S}/2$ and accounting for an additional factor of 4 as prescribed by Eq. (9), yields the following expression for the torque between two gravitomagnetic dipoles,

$$\bar{\tau}_1 = \frac{G}{c^2r^3}\bar{S}_1 \times \left[3\left(\bar{S}_2\cdot\hat{r}\right)\hat{r} - \bar{S}_2\right], \quad (12)$$

where $\bar{r}$ is the distance vector between them. The precession rate $\Omega_{LT}$, where $LT$ stands for Lense–Thirring, is defined through

$$\bar{\tau}_1 = \bar{\Omega}_{LT} \times \bar{S}_1, \quad (13)$$

so that for object 1

$$\bar{\Omega}_{LT} = -\frac{G}{c^2r^3}\left[3\left(\bar{S}_2\cdot\hat{r}\right)\hat{r} - \bar{S}_2\right]. \quad (14)$$

The measurement of such a precession rate was the central objective of the Gravity Probe B satellite experiment.[12] In this setup, the rotating Earth serves as object 2, while a set of ultra-precise gyroscopes onboard the satellite functioned as spinning test masses. The satellite was placed in a nearly polar orbit, ensuring that its trajectory remained largely aligned with the Earth's gravitomagnetic field lines. An estimate of the precession rate (14) may be obtained assuming the earth to be a perfect sphere with angular momentum

$$\bar{S}_2 = I\omega\hat{z} = \frac{2}{5}MR^2\omega\hat{z}. \quad (15)$$

Here, $\hat{z}$ is the Earth's rotation axis. The use of standard geophysical parameters gives $S_2 = 7.1\times 10^{33}\ \mathrm{kgm^2/s}$. When positioned at the equator, the first term in Eq. (14) vanishes and the precession rate becomes

$$\bar{\Omega}_{LT} = \frac{G}{c^2r^3}\bar{S}_2 \quad (16)$$

around the z-axis. In field theory, this is also the direction of the gravitomagnetic field. Using the altitude of 642 km for the satellite at this position, the rate of precession becomes $\bar{\Omega}_{LT} = 1.5\times 10^{-14}\ \hat{z}$ rad/s.

But as evident from Eq. (14), the Lense–Thirring precession rate depends on the satellite's position—specifically its latitude—and varies accordingly over the course of an orbit. Accurately determining the total accumulated precession per satellite revolution therefore requires a detailed numerical integration along the orbital path, as well as accounting for the not perfect spherical shape of the Earth.

The Gravity Probe B project collected data for nearly one year, corresponding to just over five complete satellite orbits. In 2011, the team reported a measurement of the accumulated precession that was consistent with the predictions of

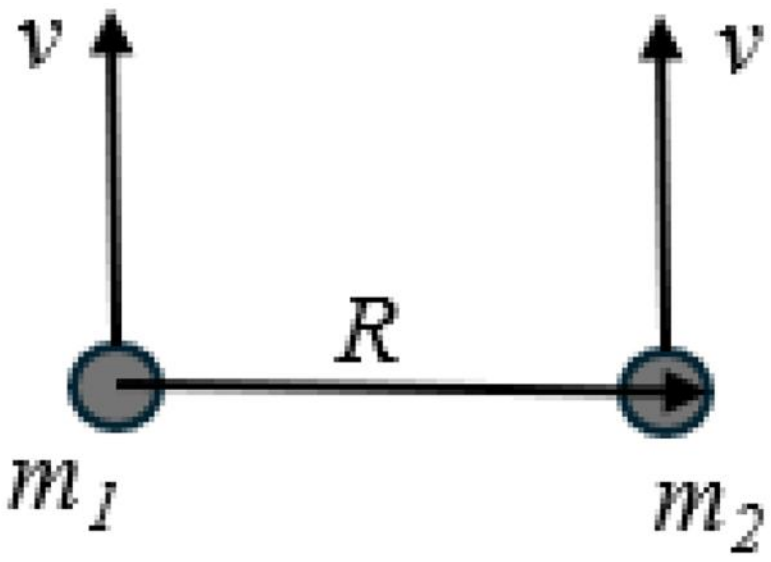

Fig. 2. Two objects in parallel motion.

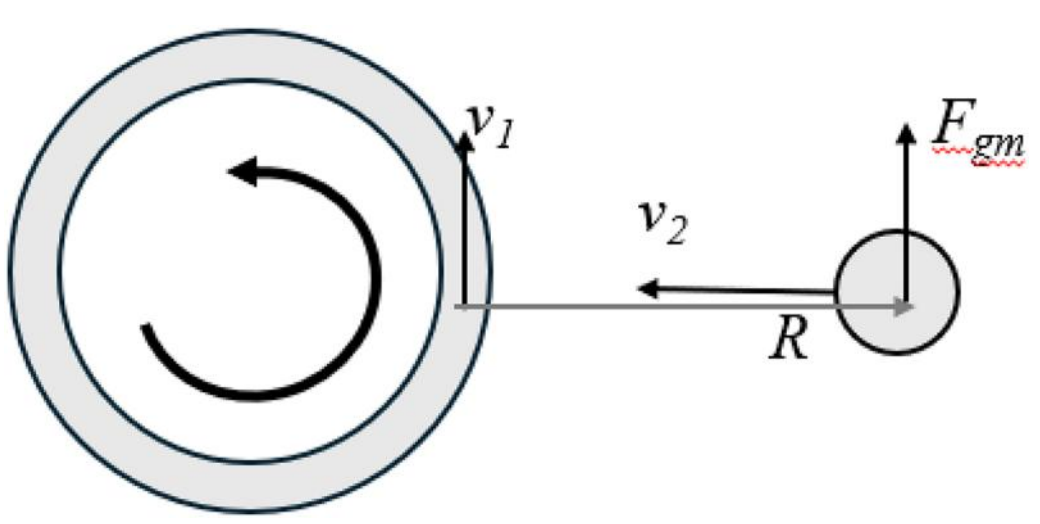

Fig. 3. Object 2 is approaching a circular mass current.

General Relativity, based on Eq. (14), with an uncertainty of 15%.

### C. Perihelion shift

The perihelion shift of Mercury's orbit stands as one of the earliest and most significant empirical confirmations of general relativity. Einstein first calculated this effect based on the Schwarzschild metric. The equation of motion was obtained in the weak-field and low-speed limit.[13] However, to reproduce this result within our framework, the Sun cannot be treated as a static mass, as assumed in the Schwarzschild solution. Instead, it must be recognized as orbiting around the barycenter of the system. This motion implies that the Sun experiences a centripetal acceleration. In the force-based interpretation, the perihelion shift is therefore associated with the acceleration-dependent component of the gravitational force. Indeed, previous studies have shown that general relativity predicts an additional radial component to the gravitational force, effectively enhancing the outward pull on Mercury.[14] This GR force will now be derived.

Within our framework, the relevant expression for the force on Mercury (object 2) due to the Sun (object 1) simplifies to

$$\bar{f}_{1\to 2} = -\frac{Gm_2m_1}{R^2}\left(1-\frac{Gm_1}{c^2R}\right)\hat{R} + \frac{4Gm_2m_1}{c^2R}\frac{d\bar{v}_1}{dt}. \tag{17}$$

Here, $\hat{R}$ denotes the unit vector pointing from the Sun to Mercury. The two gravitomagnetic terms are negligible in this context: they effectively contribute through a gravitomagnetic dipole–dipole interaction, analogous to the mechanism discussed in Sec. IV B. A direct dipole–dipole calculation confirms that its magnitude is negligible compared to the acceleration-dependent term.[15]

The acceleration in Eq. (17), $d\bar{v}_1/dt$, corresponds to the total effective acceleration of the Sun, which is inherently difficult to estimate. Nevertheless, the system can be modeled as a two-body problem, in which both bodies orbit their common center of mass and mutually influence each other through gravitational interaction. The gravitational force exerted by the Sun on Mercury is then equal in magnitude and opposite in direction to the force exerted by Mercury on the Sun, i.e., $\bar{f}_{1\to 2} = -\bar{f}_{2\to 1}$.

However, the applicability of Newton's third law in this context may be questioned. It is well known that, within classical electrodynamics, Newton's third law is generally not satisfied for interactions between free charges. This violation arises from the magnetic force, which is based on Grassmann's force law and is derived under the assumption that at least one of the interacting objects forms part of a closed conductor. In the present case, however, the magnetic contribution is absent. Consequently, Newton's third law remains valid.

Hence,

$$\begin{aligned}-\frac{Gm_2m_1}{R^2}\left(1-\frac{Gm_1}{c^2R}\right)\hat{R} &+ \frac{4Gm_2\mathrm{m}_1}{c^2R}\frac{d\bar{v}_1}{dt} \\ &= -\frac{Gm_2m_1}{R^2}\left(1-\frac{Gm_2}{c^2R}\right)\hat{R} - \frac{4Gm_2m_1}{c^2R}\frac{d\bar{v}_2}{dt}.\end{aligned} \tag{18}$$

To lowest order, an approximate circular orbit is assumed yielding $d\bar{v}_2/dt = -({v_2}^2/R)\hat{R}$.

Neglecting the term $(Gm_2m_1/R^2)(Gm_2/c^2R)$, which corresponds to the increase in the Sun's effective mass due to its interaction with Mercury, we obtain

$$\begin{aligned}\frac{4Gm_2m_1}{c^2R}\frac{d\bar{v}_1}{dt} &= -\frac{4Gm_2m_1}{c^2R}\left(-\frac{{v_2}^2}{R}\hat{R}\right) \\ &\quad -\frac{Gm_2m_1}{R^2}\left(\frac{Gm_1}{c^2R}\right)\hat{R}.\end{aligned} \tag{19}$$

The last term can be evaluated by noting that, to lowest order, the centripetal force yields $m_2v_2^2/R = Gm_2m_1/R^2 \to Gm_1/R = v_2^2$. Substituting this into the previous expression gives

$$\begin{aligned}\frac{4Gm_2m_1}{c^2R}\frac{d\bar{v}_1}{dt} &= \frac{4Gm_2m_1}{c^2R}\frac{v_2^2}{R}\hat{R} - \frac{Gm_2m_1}{R^2}\frac{v_2^2}{c^2}\hat{R} \\ &= \frac{3Gm_2m_1}{c^2R}\frac{v_2^2}{R}\hat{R}.\end{aligned} \tag{20}$$

The total force on Mercury then becomes

$$\bar{f}_{1\to 2} = -\frac{Gm_2m_1}{R^2}\left(1-3\frac{{v_2}^2}{c^2}\right)\hat{R}, \tag{21}$$

which is in agreement with the result in Ref. 14. There, the perihelion shift arising from this additional outward correction to the gravitational force is calculated and shown to agree with observational data.

Note that two distinct approaches appear to underlie the analyses. Einstein's original calculation assumes a static Sun, whereas our approach necessarily accounts for the Sun's orbital motion. Nevertheless, the two are connected via the equivalence principle, which states that the spacetime curvature produced by a static mass such as the Sun is locally indistinguishable from the effects of acceleration in flat spacetime.

A second remark concerns the role of the increase in Mercury's inertial mass due to its gravitational interaction energy with the Sun, represented by the second term in Eq. (17), i.e., the term $(Gm_2m_1/R^2\, Gm_1/c^2R)\hat{R}$. This correction reduces the predicted perihelion shift by 25%, which is a crucial contribution to reach experimental agreement. Notably, it is only through the present interaction-based analysis that this specific contribution can be explicitly identified and estimated.

### D. Origin of inertia

The acceleration-dependent term in Eq. (9) also offers an avenue for exploring the origin of inertia, and more specifically, its connection to gravitational mass as expressed in the equivalence principle. This connection was highlighted by Einstein, who emphasized that this inductive term produces a gravitational force on object 2 acting in the same direction as the acceleration of object 1.[1] He further explained that this interaction is *inductive* in nature, drawing a clear analogy to phenomena observed in electrodynamics.

Indeed, Maxwell noted that inductance can be regarded as a form of electromagnetic inertia. However, in electrodynamics, the inductive force opposes the externally applied acceleration when the interacting charges have the same sign, leading to Lenz's law. The inertial effect arises because all conduction electrons undergo the same acceleration, such that any individual electron contributes to an effective inertial response of the others through their mutual interactions, as captured by the acceleration-dependent term in Eq. (1). This phenomenon manifests itself in electromagnetic induction under time-varying currents and constitutes the underlying origin of effects such as self-inductance.

This type of collective behavior does not arise in gravitation, as there is no analogous internal process among constituents. However, the key point is that the gravitational inductive interaction acts in the *same* direction as the externally applied acceleration. This feature is central to identifying the origin of inertia. When an object is accelerated, it influences surrounding masses, effectively dragging them along. This mutual interaction gives rise to what appears as inertial resistance. Accordingly, the inertial force can be interpreted as fictitious, i.e., frame dependent. Switching to the rest frame of the accelerated object, the surrounding masses appear to accelerate in the opposite direction and thus exert a drag force on the original object. This is the fictitious inertial force observed in the accelerated frame. In the case of rotation, the centrifugal force appears in this way, being an effect due to inertia. Consequently, acceleration and inertia are relational in this perspective, in line with Mach's principle.[16]

Many authors, including Einstein, have emphasized the importance of this dynamical perspective for understanding the origin of inertia, or more precisely, the equivalence between gravitational and inertial mass.[17] A complete analysis of this phenomenon is necessarily complex, as it involves the cumulative influence of all sources in the universe, including non-visible components such as dark matter and dark energy. Because the interaction must be treated as *retarded*, accounting for the finite speed at which gravitational effects propagate, it is essential to have knowledge of the distribution of the sources not only in the present universe but also in its past. This level of detail has become available only in the last decade, through high-precision cosmological observations from the Planck and WMAP satellite missions.[18] From these data, a standard cosmological model has been developed, called ΛCDM, which provides parametrizations utilized in inertia calculations.

Many attempts have been made to perform the complete calculation, as outlined above, by summing the contributions of all mass elements up to the cosmological event horizon; see, for example, Refs. 19 and 20. Details of these calculations lie beyond the scope of the present paper and may be found in the cited references and the works cited therein. Both analyses are conducted within the framework of Eq. (9). Although mass elements near the event horizon asymptotically approach the speed of light, the applied formula pertains to the acceleration and velocity at the source, where these quantities remain small. Nonetheless, the event horizon represents a critical boundary in the calculation, as the integration tends to diverge beyond that limit. In Ref. 19, this issue was addressed by introducing an explicit cutoff, whereas in Ref. 20, a suppression factor was employed that effectively reinstated the concept of relativistic mass from special relativity. Despite these differing regularization methods, both approaches yielded consistent results, suggesting that the outcome is not highly sensitive to the precise treatment at the horizon. While the results offer support for the equivalence principle, the presence of multiple uncertainties prevents a definitive conclusion. Rather, these analyses should be regarded as a source of motivation for further investigations.

While the acceleration-dependent term in Eq. (9) offers an appealing vectorial framework for interpreting inertia, or more precisely, for supporting the equivalence principle, its role as a fundamental physical source remains contested and warrants critical examination, particularly in pedagogical settings.[21] In his later years, Einstein himself appeared skeptical of Mach's principle, especially regarding its capacity to account for the origin of inertia within the framework of general relativity.[22] It is important, however, to distinguish between two conceptually distinct perspectives. One concerns the generation of inertial mass entirely through interaction with the mass distribution of the universe—a viewpoint Einstein seems to refer to in his later years. The other, exemplified by Sultana and Kazanas[19] and Prytz,[20] focuses on reproducing the equivalence principle via gravitational induction, without invoking the generation of mass itself.

Although general relativity provides a mechanism by which inertial mass can arise from gravitationally induced dynamics, verifying this mechanism requires an analysis of the universe in its entirety, from its origin to the present epoch. Since such complete knowledge is unavailable, and is likely to remain so, Mach's principle may never be fully testable. This raises the question of whether Mach's principle should be regarded as a genuine scientific principle or rather as a philosophically motivated heuristic.

## V. TIME DILATION IN GENERAL RELATIVITY

In the supplementary material, gravitational time dilation in general relativity is derived using special relativity together with the equivalence principle and is thus recovered within a flat spacetime approximation employing the same approximation scheme as adopted in the main article. This result is subsequently used to derive the GPS timing correction, the gravitational frequency shift, and the well-known Einstein formula for light deflection.

## VI. RADIATION FROM FIELD THEORY

Radiation is inherently a field-theoretical phenomenon, and the corresponding radiated power can be derived starting from Eq. (3). In the following, we outline the derivation of gravitational radiation power, with an emphasis on its formal analogy to the electromagnetic case, an analogy that proves particularly valuable in pedagogical contexts.

Radiation originates from acceleration, which is associated with the time derivative of the vector potential. The method closely follows that of electromagnetism: the retarded vector potential is expanded in a multipole series, retaining only the leading-order term in the far zone. The Poynting vector, representing the transported power per area, is then constructed as given in Refs. 2 and 23,

$$\overline{S} = \frac{c^2}{4\pi G}\overline{E}_g \times 4\overline{B}_g, \tag{22}$$

in analogy with electromagnetism. In our notation, the gravitoelectric field in the framework is defined as

$$\overline{E}_g = \frac{c}{4}\frac{\partial \overline{A}}{\partial t} \tag{23}$$

and the gravitomagnetic field as

$$\overline{B}_g = \frac{c}{4}\nabla \times \overline{A}. \tag{24}$$

The extra factor $1/4$ compared to EM is a consequence of Eq. (4) where the vector potential $A$ is defined with the GR factor 4, which in turn is a direct result from the GR approximation obtained by Einstein.

The retarded vector potential $\overline{A}$ at the position $\overline{r} = r\hat{n}$ and at time $t$ is given by

$$\overline{A}(\overline{r},t) = \frac{4G}{c^3 r}\int \overline{J}\left(t - \frac{r}{c} + \frac{\hat{n}\cdot\overline{r'}}{c}\right) d^3r', \tag{25}$$

where $\overline{J} = \rho\overline{v}(\overline{r}')$ is the mass current. A Taylor expansion in the small time parameter $\hat{n}\cdot\overline{r'}/c$ yields

$$\overline{A}(\overline{r},t) \approx \frac{4G}{c^3 r}\left[\int \overline{J}\left(t - \frac{r}{c}\right) d^3r' + \frac{1}{c}\int \hat{n}\cdot\overline{r'}\frac{\partial\overline{J}}{\partial t} d^3r'\right]. \tag{26}$$

The first term corresponds to the dipole current and will vanish upon taking its time derivative due to momentum conservation at the source.

The second term may be expressed through the mass quadrupole momentum tensor, defined as

$$Q_j = \int \rho(\overline{r}',t) x_i' x_j' d^3r' \tag{27}$$

so that[24]

$$\int \hat{n}\cdot\overline{r'}\frac{\partial\overline{J}}{\partial t} d^3r' = \frac{1}{2}\hat{n}_j \frac{d^2 Q_{ij}}{dt^2}, \tag{28}$$

where repeated indices imply summation. Here, one term containing the gravitomagnetic dipole moment has been neglected since it will not contribute to radiation due to conservation of angular momentum at the source. The leading-order term in the multipole expansion is therefore the quadrupole term and the vector potential becomes

$$\overline{A}(\overline{r},t) \approx \frac{2G}{c^4 r}\frac{d^2 Q_{ij}}{dt^2}\hat{n}_j. \tag{29}$$

From here, we can follow the steps in Ref. 25. The gravitomagnetic field is obtained from Eq. (24),

$$\overline{B}_g(\overline{r},t) = \frac{G}{2c^4 r}\dddot{\overline{Q}}\left(\overline{r}, t - \frac{r}{c}\right) \times \hat{n}, \tag{30}$$

where

$$\dddot{\overline{Q}} = \dddot{Q}_{ij}\hat{n}_j \tag{31}$$

and the overdot denotes time derivative. The far zone gravitoelectric field becomes

$$\overline{E}_g(\overline{r},t) = \left(\frac{G}{2c^3 r}\dddot{\overline{Q}}\left(\overline{r}, t - \frac{r}{c}\right) \times \hat{n}\right) \times \hat{n}, \tag{32}$$

being perpendicular to $\overline{B}_g$ and enhanced by a factor $c$. Inserting this in the gravitomagnetic Poynting vector (22), the power is obtained after some tedious but straightforward calculations (Ref. 25, Sec. 2.3),

$$P = \frac{G}{5c^5}\dddot{Q}_{ij}\dddot{Q}_{ij}, \tag{33}$$

which was first obtained by Einstein in 1918,[26] working in the weak-field and low-speed approximation.

In 1975, this formula was successfully applied to the binary pulsar system PSR B1913 + 16, whose rate of orbital energy loss had been measured with high precision. The observed decay matched the GR prediction to within 0.2%.[27] A pedagogically adopted calculation of this energy loss is presented in Ref. 28.

Compare this formula to the corresponding electric quadrupole radiation

$$P = \frac{1}{4\pi\varepsilon_0 20c^5}\dddot{Q}_{ij}\dddot{Q}_{ij}, \tag{34}$$

where $Q_{ij}$ now stands for charge quadrupole moment.[25,29] The two formulas differ only by a factor 4 and the replacement $G \leftrightarrow 1/4\pi\varepsilon_0$, which might be a useful notification in a teaching context. This correspondence is further discussed in Ref. 29.

Similar derivations of GR radiation can be found in Refs. 28 and 30.

## VII. GR PHENOMENA BEYOND OUR APPROXIMATION

The phenomena addressed so far in this report can all be explained within the framework of linearized low-speed general relativity and are therefore well described in terms of a flat spacetime approximation. But there are several observations which go beyond this approximation:

- the detection of gravitational wave signals from merging black holes and neutron stars (e.g., GW150914),[31]
- the inference of the innermost stable circular orbit (ISCO) and accretion disk structure from x-ray spectra around black holes (e.g., Cygnus X-1, Sgr A*),[32]
- the observation of highly redshifted iron K$\alpha$ emission lines in the x-ray spectra of compact objects,[33]
- the occurrence of strong gravitational lensing phenomena such as Einstein rings, multiple quasar images, and lensing time delays,[34]
- the precise timing of binary pulsars (e.g., the Hulse–Taylor pulsar), which reveals orbital decay consistent with gravitational wave emission,[27,35]
- the established cosmological model ΛCDM being based on full General Relativity through the Friedmann–Lemaître–Robertson–Walker metric.[36]

All of these phenomena occur in regimes characterized by strong gravitational fields and therefore extend beyond the validity of linear approximations. Notably, each has been quantitatively confirmed by general relativity through either numerical nonlinear solutions or higher-order post-Newtonian calculations.

## VIII. CONCLUSIONS

We reformulated Einstein's linearized, low-speed vector approximation of GR in a manner directly analogous to the field-free interaction picture of classical electrodynamics. This approach makes it possible to identify and interpret the kinematical corrections shared between the two theories, enabling an intuitive understanding of GR without requiring mathematical tools beyond those commonly used in undergraduate electromagnetism.

Within this formulation, we demonstrated that the core predictions of general relativity are accessible and transparent. As such, our approach illustrates the equivalence between the conventional geometrical interpretation based on curved spacetime and a force-based perspective in flat spacetime. These two formulations are fundamentally linked through Noether's theorem,[37] which connects symmetries to conserved quantities.

Noether's theorem shows that the conservation of energy and momentum is directly linked to the uniformity of spacetime. In a perfectly uniform spacetime, energy and momentum are conserved in the usual sense. In general relativity, spacetime can be curved and therefore not uniform from place to place. As a result, the momentum of an individual object may change as it moves through spacetime, which, in a force-based description, is interpreted as the action of a gravitational force, even though the total energy and momentum of the complete system remain conserved.

Notably, Einstein himself emphasized that the geometrical interpretation of general relativity should not be regarded as fundamental. In his own words from 1926:[38]

> "It is wrong to think that 'geometrization' is something essential. It is only a kind of crutch (Eselsbrücke) for the finding of numerical laws. Whether one links 'geometrical' intuitions with a theory is a private matter."

Einstein's own reflections serve as a compelling motivation to explore alternative formulations of general relativity that enhance its conceptual accessibility. In this report, we aimed to provide such a perspective.

## SUPPLEMENTARY MATERIAL

Please click on this link to access the supplementary material, where gravitational time dilation in general relativity is derived using special relativity together with the equivalence principle and is thus recovered within a flat spacetime approximation employing the same approximation scheme as adopted in the main article. Print readers can see the supplementary material at https://doi.org/10.60893/figshare.ajp.c.8464410.

## ACKNOWLEDGMENTS

I would like to express my sincere gratitude to Dr. Sverker Edvarsson for valuable discussions, and to the four reviewers of the manuscript for their insightful and constructive comments.

[a]Electronic mail: Kjell.Prytz@mdu.se, ORCID: 0000-0001-8339-6714.

[1]A. Einstein, *The Meaning of Relativity*, 1st ed. (Methuen & Co. Ltd, London, 1922); 6th ed. (Chapman and Hall, London, 1956). https://archive.org/details/meaningofrelativ00eins_0/page/112/mode/2up

[2]B. Mashhoon, "Gravitomagnetism: A brief review," arXiv:gr-qc/0311030 (2008); Appears also in: *The Measurement of Gravitomagnetism: A Challenging Enterprise*, edited by L. Iorio (Nova Science, New York, 2007), pp. 29–39.

[3]A. K. T. Assis, *Weber's Electrodynamics* (Kluwer Academic Publishers, Boston, 1994); K. Prytz, *Electrodynamics—The Field Free Approach* (Springer, Cham, 2015).

[4]A. M. Ampère, "Mémoire sur la théorie mathématique des phénomènes électrodynamiques uniquement déduite de l'expérience," Mém. Acad. Sci. Inst. Fr. **6**, 175–388 (1823).

[5]M. Faraday, "Experimental researches in electricity. On the induction of electric currents," Philos. Trans. R. Soc. London **122**, 125–162 (1832); J. Henry, "On the production of currents and sparks of electricity from magnetism," Am. J. Sci. Arts **22**, 408–418 (1832).

[6]W. Weber, "On the measurement of electrodynamic forces," Ann. Phys. **73**, 193–240 (1848); C. F. Gauss, "Zur mathematischen theorie der elektrodynamischen wirkungen," in *Werke* (Königlichen Gesellschaft der Wissenschaften, Göttingen, 1867), Vol. 5; W. Weber, "Elektrodynamische maassbestimmungen: Über ein allgemeines grundgesetz der elektrischen wirkung," in *Werke* (Julius Springer, Berlin, 1893), p. 25.

[7]P. Moon and D. E. Spencer, "A new electrodynamics," J. Franklin Inst. **257**(5), 369–382 (1954); P. Moon and D. E. Spencer, "Electromagnetism without magnetism—An historical sketch," Am. J. Phys. **22**, 120–123 (1954).

[8]P. Lorrain and D. Corson, *Electromagnetic Fields and Waves*, 3rd ed. (W. H. Freeman, New York, 1988); L. Page, "A derivation of the fundamental relations of electrodynamics from those of electrostatics," Am. J. Sci. **34** (199), 57–68 (1912).

[9]J. Lense and H. Thirring, "Über den einfluss der eigenrotation der zentralkörper auf die bewegung der planeten und monde nach der einsteinschen relativitätstheorie," Physikalische Z **19**, 156–163 (1918).

[10]B. Mashhoon, "On the gravitational analogue of Larmor's theorem," Phys. Lett. A **173**, 347–354 (1993).

[11]K. Prytz, *Electrodynamics—The Field Free Approach* (Springer, Cham, 2015), Chap. 7.

[12]C. W. F. Everitt *et al*, "Gravity Probe B: Final results of a space experiment to test general relativity," Phys. Rev. Lett. **106**(22), 221101 (2011); I. Ciufolini and C. E. Pavlis, "A confirmation of the general relativistic prediction of the Lense–Thirring effect," Nature **431**(7011), 958–960 (2004).

[13]A. Einstein, *The Collected Papers of Albert Einstein* (Princeton U. P., Princeton, NJ, 1996), Vol. 6, p. 838; see also https://www.researchgate.net/publication/228923053_Einstein%27s_PaperExplanation_of_the_Perihelion_Motion_of_Mercury_from_General_Relativity_Theory for "Einstein's Paper: Explanation of the Perihelion Motion of Mercury from General Relativity Theory."

[14]B. Davies, "Elementary theory of perihelion precession," Am. J. Phys. **51**(10), 909–911 (1983); S. Edvardsson, "Relativistic gravitational force," Celest. Mech. Dyn. Astron. **135**(3), 25 (2023); G. S. Adkins and J. McDonnell, "Orbital precession due to central-force perturbations," Phys. Rev. D **75**(8), 082001 (2007); J. Bootello, "Relativistic perihelion precession of orbits of celestial bodies," Int. J. Astron. Astrophys. **2**, 249–255 (2012).

[15]I. Haranas, O. Ragos, and I. Gkigkitzis, "The Lense–Thirring effect in the anomalistic period of celestial bodies," Am. J. Space Sci. **1**(2), 46–53 (2013).

[16]E. Mach, *The Science of Mechanics* (The Open Court Publishing Co., Chicago/London, 1919).

[17]K. Nordtvedt, "Gravitomagnetism in local inertial frames," Int. J. Theor. Phys. **27**, 1395–1404 (1988); D. W. Sciama, "On the origin of inertia," Mon. Not. R. Astron. Soc. **113**(1), 34–42 (1953); A. P. French, *Newtonian Mechanics*, The MIT Introductory Physics Series (W. W. Norton, New York, 1971); A. K. T. Assis, "Weber's electrodynamics and the origin of inertia," Found. Phys. Lett. **26**, 271–283 (1996); D. J. Raine, "Mach's principle," Rep. Prog. Phys. **44**, 1151–1187 (1981); L. R. Signore, "Mach's principle and the origin of inertia: A modern perspective," Il Nuovo Cimento B **111**, 1087–1103 (1996).

[18]N. Jarosik *et al*., "Seven-year Wilkinson Microwave Anisotropy Probe (WMAP) observations: Sky maps, systematic errors, and basic results," Astrophys. J., Suppl. Ser. **192**(2), 14 (2011); P. A. R. Ade *et al*., "*Planck 2013* results. I. Overview of products and scientific results," Astron. Astrophys. **571**, A16 (2014).

[19]J. Sultana and D. Kazanas, "The problem of inertia in Friedmann universes," Int. J. Mod. Phys. D **20**(07), 1205–1214 (2011).

[20]K. Prytz, "Sources of inertia in an expanding universe," Open Phys. **13**(1), 130–134 (2015).

[21]L. L. Williams and N. Inan, "Frame-dragging and the origin of inertia: New developments," New J. Phys. **23**, 053019 (2021); D. Bini, C. Cherubini, R. T. Jantzen, and B. Mashhoon, "Gravitomagnetism and relative observer clock effects," Classical Quantum Gravity **25**, 225014

(2008); C. Brans, “Mach’s principle and a relativistic theory of gravitation,” Phys. Rev. **125**, 2194–2201 (1962).
[22]G. F. R. Ellis and R. Penrose, “Dennis William Sciama. 18 November 1926—19 December 1999,” Biogr. Mem. Fellows R. Soc. **56**, 401–422 (2010).; see also: https://vtechworks.lib.vt.edu/server/api/core/bitstreams/dc88d4ee-95f8-444f-989f-78d0a9c51817/content for “We Shall Have to Make the Best of It’: The Conversion of Dennis Sciama”
[23]Gravitoelectromagnetism, Wikipedia, see https://en.wikipedia.org/wiki/Gravitoelectromagnetism for “The Free Encyclopedia” (accessed September 4, 2025).
[24]K. K. Likharev, see https://phys.libretexts.org/Bookshelves/Electricity_and_Magnetism/Essential_Graduate_Physics_-_Classical_Electrodynamics_(Likharev) for “Essential Graduate Physics: Classical Electrodynamics, LibreTexts Physics” (accessed September 4, 2025).
[25]G. C. Dorsch and L. E. A. Porto, “An introduction to gravitational waves through electrodynamics: A quadrupole comparison,” Eur. J. Phys. **43**(2), 025602 (2022).
[26]A. Einstein, “Über gravitationswellen [On gravitational waves],” Sitzungsber. K. Preuss. Akad. Wiss., 154–167 (1918); S. Eddington, “The propagation of gravitational waves,” Proc. R. Soc. A **102**, 268–282 (1922).
[27]R. A. Hulse and J. H. Taylor, “Discovery of a pulsar in a binary system” Astrophys. J. Lett. **195**, L51–L53 (1975); J. H. Taylor, L. A. Fowler, and P. M. McCulloch, “Measurements of general relativistic effects in the binary pulsar PSR 1913 + 16,” Nature **277**(5696), 437–440 (1979).
[28]R. C. Hilborn, “Gravitational waves without general relativity redux,” Am. J. Phys. **92**(10), 780–785 (2024).
[29]P. Christillin and L. Barattini, “Gravitomagnetic forces and quadrupole gravitational radiation from special relativity,” arXiv:1205.3514 [gr-qc] (2012).
[30]I. A. Arbab, “On the gravitational radiation of gravitating objects,” Astrophys. Space Sci. **323**(2), 181–184 (2009).
[31]B. P. Abbott *et al.*, “Observation of gravitational waves from a binary black hole merger,” Phys. Rev. Lett. **116**(6), 061102 (2016).
[32]J. M. Miller, “Relativistic X-ray lines from the inner accretion disks around black holes,” Annu. Rev. Astron. Astrophys. **45**(1), 441–479 (2007).
[33]A. C. Fabian, M. J. Rees, L. Stella, and N. E. White, “X-ray fluorescence from the inner disc in Cygnus X-1,” Mon. Not. R. Astron. Soc. **238**(3), 729–736 (1989).
[34]S. Refsdal and H. Bondi, “On the possibility of determining Hubble’s parameter and the masses of galaxies from the gravitational lens effect,” Mon. Not. R. Astron. Soc. **128**(4), 295–306 (1964) [First proposal of time-delay lensing]; T. Treu and P. J. Marshall, “Time delay cosmography,” Astron. Astrophys. Rev. **24**, 11 (2016).
[35]J. M. Weisberg and Y. Huang, “Relativistic measurements from timing the binary pulsar PSR B1913 + 16,” Astrophys. J. **829**(1), 55 (2016).
[36]P. J. E. Peebles and B. Ratra, “The cosmological constant and dark energy,” Rev. Mod. Phys. **75**(2), 559–606 (2003).
[37]E. Noether, “Invariante variationsprobleme [Invariant variation problems],” Nach. Ges. Wiss. Göttingen, Math.-Phys. Klasse, 235–257 (1918); Transp. Theory Stat. Phys. **1**, 186–207 (1971).
[38]D. Lehmkuhl, “Why Einstein did not believe that general relativity geometrizes gravity,” Stud. History Philos. Sci., Part B **46**, 316–326 (2014).

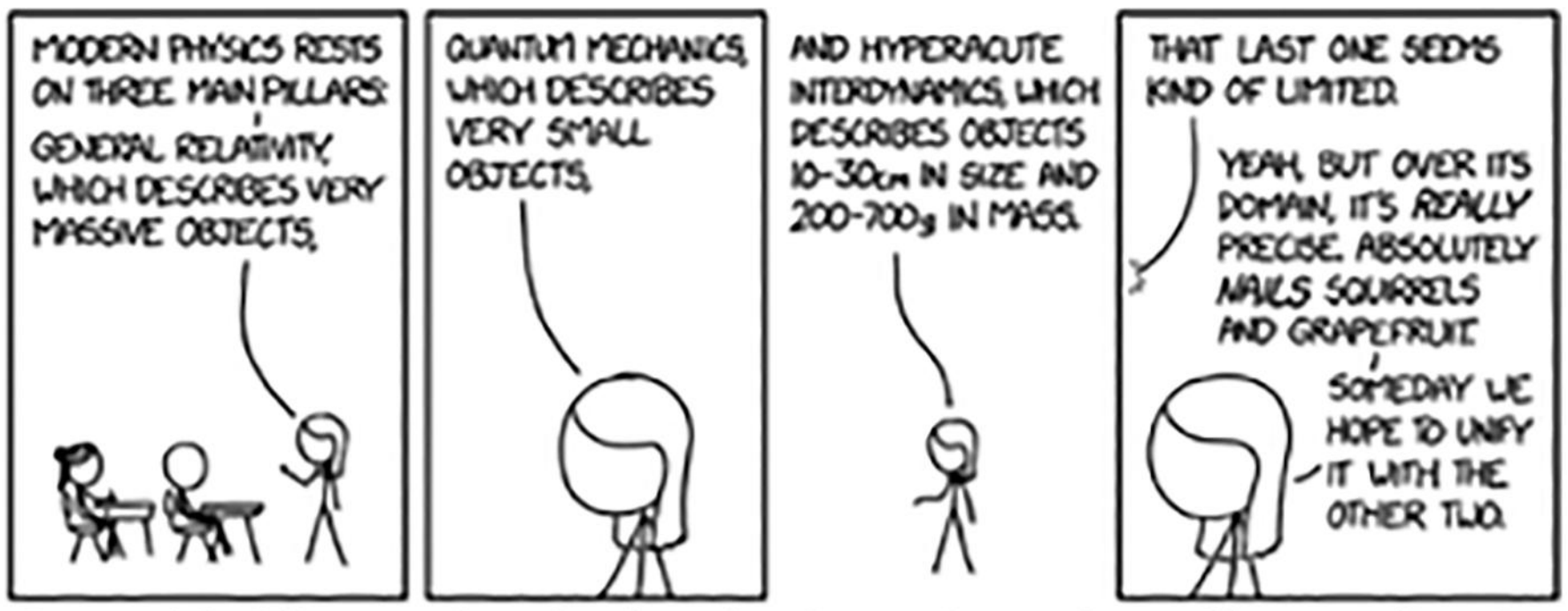

Our models fall apart where the three theories overlap; we're unable to predict what happens when a nanometer-sized squirrel eats a grapefruit with the mass of the sun. (Source: https://xkcd.com/3178/)